\documentclass[aps,prl,superscriptaddress,amsmath,twocolumn,showpacs]{revtex4-1}

\usepackage{hangcaption,hhline,psfrag,rotating,amssymb}
\usepackage[hang,nooneline]{subfigure}
\usepackage{dcolumn}
\usepackage{bm}
\usepackage[pdftex]{color}
\usepackage[sort&compress]{natbib}
\usepackage[colorlinks=true,linkcolor=blue,filecolor=blue,urlcolor=blue,citecolor=blue,pdftex=true,plainpages=false]{hyperref}

\begin{document}
\author{Jon A.~Bailey}
\affiliation{Department of Physics and Astronomy, Seoul National University, Seoul, South Korea}

\author{A.~Bazavov}
\affiliation{Physics Department, Brookhaven National Laboratory, Upton, NY, USA}

\author{C.~Bernard}
\affiliation{Department of Physics, Washington University, St.~Louis, Missouri, USA}

\author{C.M.~Bouchard}
\affiliation{Department of Physics, The Ohio State University, Columbus, Ohio, USA}

\author{C.~DeTar}
\affiliation{Physics Department, University of Utah, Salt Lake City, Utah, USA}

\author{Daping~Du}
\email{ddu@illinois.edu}
\affiliation{Physics Department, University of Illinois, Urbana, Illinois, USA}

\author{A.X.~El-Khadra}
\affiliation{Physics Department, University of Illinois, Urbana, Illinois, USA}

\author{J.~Foley}
\affiliation{Physics Department, University of Utah, Salt Lake City, Utah, USA}

\author{E.D.~Freeland}
\affiliation{Department of Physics, Benedictine University, Lisle, Illinois, USA}

\author{E.~G\'amiz}
\affiliation{CAFPE and Departamento de F\`isica Te\'orica y del Cosmos, Universidad de Granada,
Granada, Spain}

\author{Steven~Gottlieb}
\affiliation{Department of Physics, Indiana University, Bloomington, Indiana, USA}

\author{U.M.~Heller}
\affiliation{American Physical Society, Ridge, New York, USA}

\author{Jongjeong Kim}
\affiliation{Department of Physics, University of Arizona, Tucson, Arizona, USA}

\author{A.S.~Kronfeld}
\affiliation{Fermi National Accelerator Laboratory, Batavia, Illinois, USA}

\author{J.~Laiho}
\affiliation{SUPA, School of Physics and Astronomy, University of Glasgow, Glasgow, UK}

\author{L.~Levkova}
\affiliation{Physics Department, University of Utah, Salt Lake City, Utah, USA}

\author{P.B.~Mackenzie}
\affiliation{Fermi National Accelerator Laboratory, Batavia, Illinois, USA}

\author{Y.~Meurice}
\affiliation{Department of Physics and Astronomy, University of Iowa, Iowa City, Iowa, USA}

\author{E.T.~Neil}
\affiliation{Fermi National Accelerator Laboratory, Batavia, Illinois, USA}

\author{M.B.~Oktay}
\affiliation{Physics Department, University of Utah, Salt Lake City, Utah, USA}

\author{Si-Wei Qiu}
\affiliation{Physics Department, University of Utah, Salt Lake City, Utah, USA}

\author{J.N.~Simone}
\affiliation{Fermi National Accelerator Laboratory, Batavia, Illinois, USA}

\author{R.~Sugar}
\affiliation{Department of Physics, University of California, Santa Barbara, California, USA}

\author{D.~Toussaint}
\affiliation{Department of Physics, University of Arizona, Tucson, Arizona, USA}

\author{R.S.~Van~de~Water}
\email{ruthv@bnl.gov}
\affiliation{Physics Department, Brookhaven National Laboratory, Upton, NY, USA}

\author{Ran Zhou}
\affiliation{Department of Physics, Indiana University, Bloomington, Indiana, USA}

\collaboration{Fermilab Lattice and MILC Collaborations}
\noaffiliation

\preprint{FERMILAB-PUB-12/333-T }

\title{\boldmath Refining new-physics searches in $B \to D \tau \nu$ decay with lattice QCD}

\date{\today}

\begin{abstract}
The semileptonic decay channel $B\to D \tau \nu$ is sensitive to the presence of a scalar current, such as
that mediated by a charged-Higgs boson.
Recently the BaBar experiment reported the first observation of the exclusive semileptonic decay 
$B\to D\tau^-\overline{\nu}$, finding an approximately 2$\sigma$ disagreement with the
Standard-Model prediction for the ratio $R(D)=\text{BR}(B\to D\tau\nu)/\text{BR}(B\to D\ell\nu)$, where
$\ell=e,\mu$.
We compute this ratio of branching fractions using hadronic form factors computed in unquenched lattice QCD
and obtain $R(D) = 0.316(12)(7)$, where the errors are statistical and total systematic, respectively.
This result is the first Standard-Model calculation of $R(D)$ from {\it ab initio} full QCD.
Its error is smaller than that of previous estimates, primarily due to the reduced uncertainty 
in the scalar form factor~$f_0(q^2)$.
Our determination of $R(D)$ is approximately 1$\sigma$ higher than previous estimates and, thus,
reduces the tension with experiment.  
We also compute $R(D)$ in models with electrically charged scalar exchange, such as the type~II two-Higgs
doublet model.
Once again, our result is consistent with, but approximately 1$\sigma$ higher than, previous estimates for phenomenologically relevant values of the scalar coupling in the type~II model.
As a byproduct of our calculation, we also present the Standard-Model prediction for the longitudinal
polarization ratio $P_L (D)= 0.325(4)(3)$.
\end{abstract}


\maketitle


{\it Motivation}.
-- The third generation of quarks and leptons may be particularly sensitive to new physics associated with
electroweak symmetry breaking due to their larger masses.
For example, in the minimal supersymmetric extension of the Standard Model, charged-Higgs contributions to
tauonic $B$ decays can be enhanced if $\tan\beta$ is large.
Thus the semileptonic decay $B \to D \tau \nu$ is a promising new-physics search
channel~\cite{Grzadkowski:1992qj,Tanaka:1994ay,Garisto:1994vz,Tsai:1996ps,Wu:1997uaa,Kiers:1997zt,%
Miki:2002nz,Itoh:2004ye,Chen:2006nua,Nierste:2008qe,Kamenik:2008tj,Tanaka:2010se,Fajfer:2012vx}.

The BaBar experiment recently measured the ratios 
$R(D^{(*)})=\text{BR}(B \to D^{(*)} \tau \nu)/\text{BR}(B\to D^{(*)}\ell\nu)$, where $\ell = e,\mu$, and 
reported excesses in both channels which, when combined, disagree with the Standard Model by 
$3.4\sigma$~\cite{:2012xj}.
BaBar also interpreted these measurements in terms of the type-II two-Higgs-doublet model (2HDM~II) and claimed
to exclude the theory at 99.8\% confidence-level.
In this work, we update the prediction for $R(D)$ in the Standard-Model and in new-physics theories with a
scalar current (such as the 2HDM~II) using unquenched lattice-QCD calculations of the $B \to D \ell \nu$ form
factors $f_0(q^2)$ and $f_+(q^2)$ by the Fermilab Lattice and MILC collaborations~\cite{Bailey:2012rr}.
This is the first determination of $R(D)$ from {\it ab initio} full QCD.

With lepton helicity defined in the rest frame  of the virtual $W$~boson, the general expressions for the 
differential rates for semileptonic $B \to D \ell \nu$ decay are given by
\begin{widetext}
\begin{eqnarray}
	\frac{d \Gamma_- }{dq^2} & = &  \frac{1}{24 \pi^3} 
		\left(1-\frac{m_\ell^2}{q^2}\right)^2 |\bm{p}_D|^3 
		\left|{G}_V^{\ell cb} f_+(q^2) - \frac{m_\ell}{M_B} G_T^{\ell cb} f_2(q^2)\right|^2 \,,
    \label{eq:dGammaMinus}
    \\[1em]
	\frac{d \Gamma_+ }{dq^2} & = &  \frac{1}{16\pi^3} 
		\left(1-\frac{m_\ell^2}{q^2}\right)^2 \frac{|\bm{p}_D|}{q^2} 
    \left\{\frac{1}{3} |\bm{p}_D|^2
		\left|m_\ell{G}_V^{\ell cb} f_+(q^2) - \frac{q^2}{M_B}G_T^{\ell cb} f_2(q^2)\right|^2
	\right. \label{eq:dGammaPlus} \\ & + & \left.
        \frac{\left(M_B^2-M_D^2\right)^2}{4M_B^2} 
        \left|\left(m_\ell{G}_V^{\ell cb} - \frac{q^2}{m_b-m_c} G_S^{\ell cb} \right) f_0(q^2)\right|^2 \right\} \,, \nonumber 
\end{eqnarray}
\end{widetext}
where the subscript denotes helicity, and $q = (p_\ell + p_{\nu})$ is the momentum carried by the charged
lepton-neutrino pair.
The total semileptonic width is the sum of the partial widths, $\Gamma_\text{tot} = (\Gamma_+ + \Gamma_-)$.
At tree-level of the Standard Model electroweak interaction, the scalar- and tensor-exchange couplings are
$G_S = G_T = 0$, while the vector coupling is ${G}_{V}^{\ell ij} = G_FV_{ij} $.
In the infinite heavy-quark-mass limit, the form factors $f_+(q^2)$ and $f_2(q^2)$ are related via
\begin{equation}
    f_2(q^2) = - f_+(q^2) - \frac{M_B^2-M_D^2}{q^2}\left[f_+(q^2)-f_0(q^2)\right] \,.
    \label{eq:f2static}
\end{equation}
The form factor $f_2(q^2)$ is only relevant for theories with tensor currents, however, which we do not
consider here.

Because the Standard-Model positive-helicity contribution to semileptonic $B \to D$ decay is proportional to
the lepton mass-squared, it can be neglected for the light leptons $\ell = e,\mu$; thus experimental
measurements of $B \to D \ell \nu$ decays are sensitive only to the vector form factor $f_+(q^2)$.
On the other hand, the differential rate for $B\to D\tau\nu$ is sensitive also to the positive-helicity 
contribution and, hence, $f_0(q^2)$.
Existing Standard-Model estimates of $d\Gamma_-/dq^2$ for $B\to D\tau\nu$ have relied on the kinematic 
constraint $f_0(0)=f_+(0)$, dispersive bounds on the shape~\cite{Caprini:1997mu}, relations from heavy-quark 
symmetry, and quenched lattice QCD (neglecting $u$, $d$, and $s$ quark 
loops)~\cite{deDivitiis:2007ui,deDivitiis:2007uk}. 
See Refs.~\cite{Kamenik:2008tj,Nierste:2008qe,Tanaka:2010se,Fajfer:2012vx} for details.
In this letter, we replace quenched QCD and heavy-quark estimates with a full, 2+1-flavor QCD calculation.
In particular, we determine the following ratios within the Standard Model (where $\ell = e,\mu$):
\begin{eqnarray} 
	R(D) &=& \text{BR} (B \to D \tau \nu) /  \text{BR} (B \to D \ell \nu) \label{eq:RD} \,, \\
	P_L(D) &=& \left( \Gamma_+^{B\to D \tau \nu} - \Gamma_-^{B\to D \tau \nu}  \right) /  \Gamma_{\rm tot}^{B\to D \tau \nu} \,. \label{eq:PL}
\end{eqnarray}
These quantities enable particularly clean tests of the Standard Model and probes of new physics because the
CKM matrix elements and many of the hadronic uncertainties cancel between the numerator and denominator.


{\it Lattice-QCD calculation}.
-- Here we briefly summarize the lattice-QCD calculation of the $B\to D\ell\nu$ semileptonic form factors
$f_+(q^2)$ and $f_0(q^2)$~\cite{Bailey:2012rr}.
Our calculation is based on a subset of the (2+1)-flavor ensembles generated by the MILC
Collaboration~\cite{Bazavov:2009bb}.
We use two lattice spacings $a\approx 0.12$~and 0.09~fm, and two light-quark masses at each lattice spacing
with (Goldstone) pion masses in the range 315--520~MeV; the specific numerical simulation parameters are given in Table~I of Ref.~\cite{Bailey:2012rr}.

This relatively small data set is sufficient for ratios such as those studied here and in
Ref.~\cite{Bailey:2012rr}, given the mild chiral and continuum extrapolations.
We use the Fermilab action~\cite{ElKhadra:1996mp} for the heavy quarks (bottom and charm) and use the
asqtad-improved staggered action~\cite{Bazavov:2009bb} for the light valence and sea quarks ($u,d,s$).
We minimize the systematic error due to contamination from radial excitations in 2-point and 3-point
correlation functions by employing fits including their contributions, as described in
Sec.~III of Ref.~\cite{Bailey:2012rr}. We renormalize the lattice vector current $\bar{c}\gamma^\mu b$ (and
other heavy-heavy currents) using a mostly nonperturbative method~\cite{ElKhadra:2001rv} in which we
determine the flavor-conserving normalizations nonperturbatively.
The remaining correction is close to unity and can be calculated in one-loop tadpole-improved lattice
perturbation theory~\cite{Lepage:1992xa}.
When extrapolating the lattice simulation results to the physical light-quark masses and the continuum limit,
we carefully account for the leading nonanalytic dependence on the light-quark masses at nonzero but small
momentum transfer \cite{Chow:1993hr} including the effects of lattice artifacts (generic discretization
errors and taste-symmetry breaking introduced by the staggered action)~\cite{Aubin:2005aq, Laiho:2005ue}.
The chiral-continuum extrapolation results are plotted in the left panels of Figs.~6 and~7 of
Ref.~\cite{Bailey:2012rr}; for details see Eqs.~(4.1) and~(4.2) of the same work and the surrounding text.

\begin{figure}[t]
    \includegraphics[width=\linewidth]{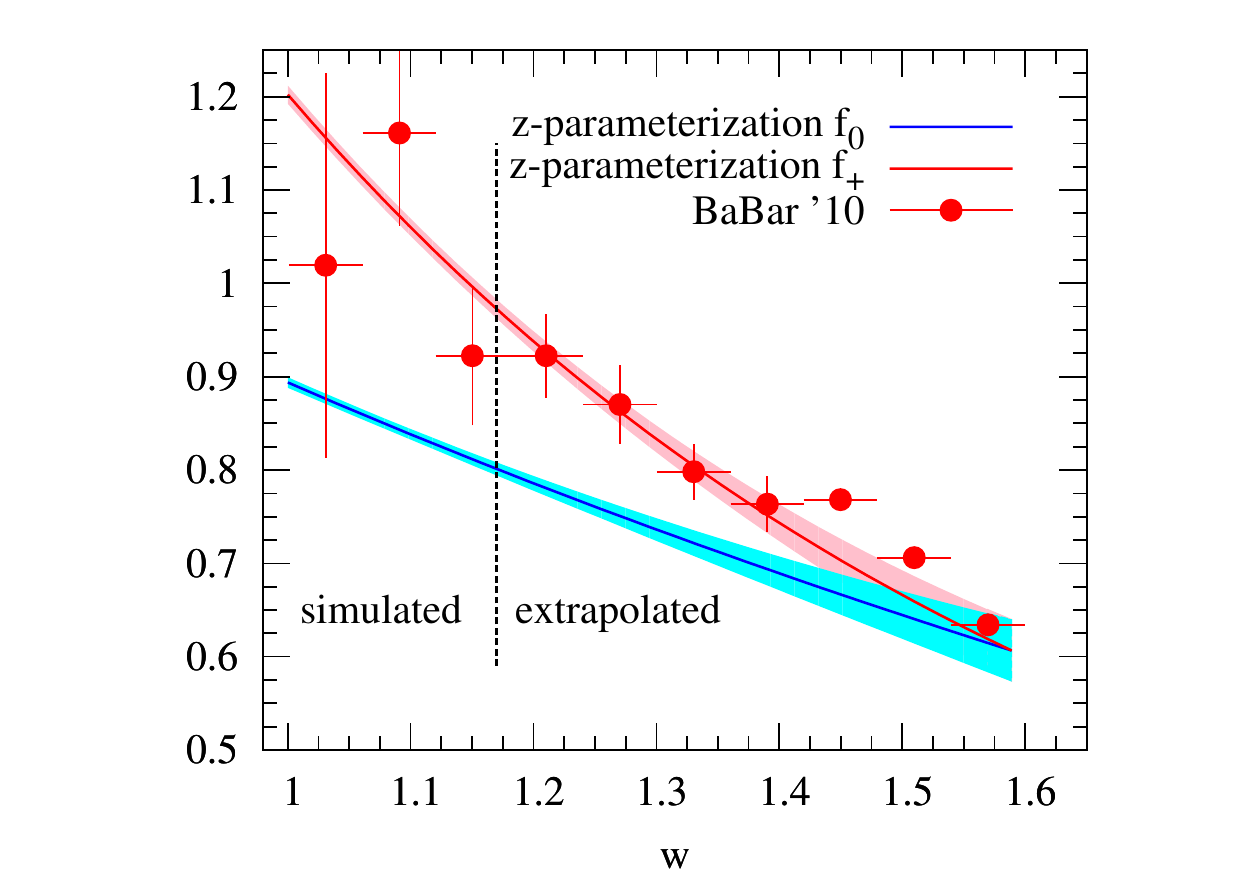}
    \caption{The form factors $f_+$ (top, red) and $f_0$ (bottom, blue) from lattice QCD.
    The range of simulated recoil values is to the left of the vertical line.
    The filled colored bands show the interpolation/extrapolation of the numerical lattice data over the full
    kinematic range using the $z$~parameterization.
    For comparison, the experimental measurement from BaBar~\cite{Aubert:2009ac} is shown as solid filled 
    circles (using $|V_{cb}|=41.4 \times 10^{-3}$~\cite{deDivitiis:2007ui,Aubert:2009ac}).}
    \label{fig:lattice_f0_fp}
\end{figure}

Figure~\ref{fig:lattice_f0_fp} shows the results for $f_+(q^2)$ and $f_0(q^2)$~\cite{Bailey:2012rr}.
The simulated data are in the range $w < 1.17$ (to the left of the dashed vertical line), where
$w=(M_B^2+M_D^2 - q^2)/(2M_BM_D)$.
In this region, we parameterize the $w$~dependence of the form factors by a quadratic expansion about $w=1$,
which works well for small recoil.
To extend the form-factor results beyond the simulated recoil values (to the right of the dashed vertical
line) we reparameterize the form factors in terms of the variable $z$~\cite{Boyd:1997kz}, and then
extrapolate to large recoil using a model-independent fit function based on general quantum-theory-principles
of analyticity and crossing-symmetry.
The functional forms used to extrapolate $f_+(q^2)$ and $f_0(q^2)$ are defined in
Eqs.~(5.1)--(5.6) of Ref.~\cite{Bailey:2012rr}.
The fits are plotted in the left panel of Fig.~9, and the results are given in the upper panel of Table~V of
the same work.
As seen in Fig.~\ref{fig:lattice_f0_fp}, our result for $f_+(q^2)$ agrees very well with experimental
measurements~\cite{Aubert:2009ac} over the full kinematic range.
This nontrivial check gives confidence in the extrapolation of $f_0(q^2)$, which cannot be obtained
experimentally and for which lattice-QCD input is crucial.
In particular, lattice-QCD uncertainties are smallest near $q^2=(M_B-M_D)^2$, so the discussion below hinges
principally on our calculation of $f_0(q^2)$ near this point, the validated $f_0(0)=f_+(0)$, and a smooth
connection between the two limits.

\begin{figure}[t]
    \includegraphics[width=\linewidth]{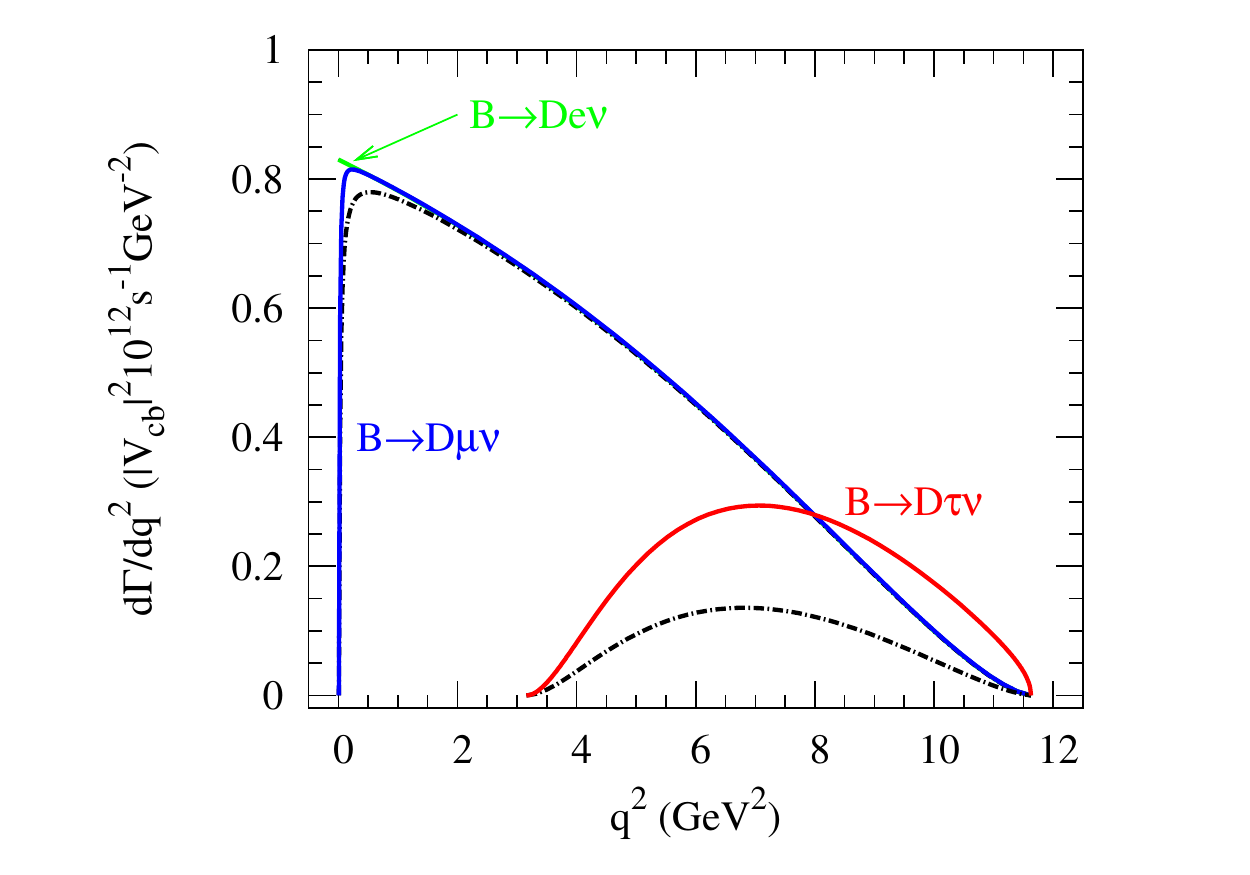}
    \caption{Differential decay rates for $B \to D e \nu$ (green), $B \to D \mu \nu$ (blue), and 
    $B \to D \tau \nu$ (red) in the Standard Model.   
    The black dot-dashed curves show the rates calculated with $f_0(q^2)=0$.} 
    \label{fig:partial_rates}
\end{figure}

We calculate the Standard-Model $B \to D\ell\nu$ partial decay rates into the three generations of leptons
using these form factors and Eqs.~(\ref{eq:dGammaMinus}) and~(\ref{eq:dGammaPlus}) with $G_S=G_T=0$, 
$G_V=G_FV^*_{cb}$.
The resulting distributions are plotted in Fig.~\ref{fig:partial_rates}.
To illustrate the role of the scalar form factor $f_0(q^2)$, we also show the rates with only the
contributions from $f_+(q^2)$.
Due to the significant helicity suppression, the differential decay rates into light leptons are
well-approximated by a single contribution from the form factor $f_+(q^2)$.
For $B\to D\tau\nu$, however, the contribution from the scalar form factor $f_0(q^2)$ comprises half of the
Standard-Model rate.

Given the lattice-QCD determinations of $f_+(q^2)$ and $f_0(q^2)$ we can obtain the Standard-Model values for
$R(D)$ and $P_L(D)$.
These are the primary results of this letter, and we now discuss the sources of systematic uncertainty.
In Ref.~\cite{Bailey:2012rr}, many statistical and several systematic errors cancelled approximately or
exactly in the ratio $f_0^{B_s\to D_s\ell\nu}/f_0^{B\to D\ell\nu}$ studied there.
Some of these do not cancel (as well) in $R(D)$ and $P_L(D)$, however, because they affect $f_+(q^2)$ and
$f_0(q^2)$ differently.

Table~\ref{tab:ErrBud} shows the error budgets for $R(D)$ and $P_L(D)$.
The statistical error in $R(D)$ is significant (3.7\%) due to the different phase-space integrations in the
numerator and denominator, whereas for $P_L(D)$ the correlated statistical fluctuations largely cancel.
For the same reason, the errors in $R(D)$ arising from the extrapolation to the physical light-quark masses
and the continuum limit ($1.4\%$) and to the full $q^2$ range ($1.5\%$), are much larger than for $P_L(D)$.
We estimate the error from the chiral-continuum extrapolation by comparing the results for fits with and
without next-to-next-to-leading order analytic terms in the chiral expansion.
We estimate the error from the $z$ extrapolation by varying the range of synthetic data used in the
$z$ fit, including an additional pole in the fit function, and including higher powers of~$z$.
The specific chiral and $z$-fit variations considered are enumerated in Table~VI of Ref.~\cite{Bailey:2012rr} and discussed in detail in the surrounding text.
The remaining sources of uncertainty in Table~\ref{tab:ErrBud} do not contribute significantly to the
quantities studied in Ref.~\cite{Bailey:2012rr}, so we describe them in greater detail below.

We determine the bare heavy-quark masses in our simulations by tuning the parameters $\kappa_b$ and
$\kappa_c$ in the heavy-quark action such that the kinetic masses of the pseudoscalar $B_s$ and $D_s$ mesons
match the experimentally-measured values~\cite{ElKhadra:1996mp}.
In practice, it is easier to work with the form factors $h_{\pm}(w)$ on the lattice, which are linear
combinations of $f_{+,0}(q^2)$~\cite{Bailey:2012rr}.
We study how the form factors $h_\pm(w)$ depend on $\kappa_{b,c}$ by re-computing the form factors on some 
ensembles at values of $\kappa_{b,c}$ slightly above and below the default ones, and extracting
the slopes with respect to $\kappa_{b,c}$.
We use these slopes to correct our results for $R(D)$ and $P_L(D)$ slightly from the simulated $\kappa$
values to the physical ones, and conservatively take the full size of the shift as the error due to
$\kappa$-tuning.

We remove the leading taste-breaking light-quark discretization errors in the form factors with the
chiral-continuum extrapolation, and estimate the remaining discretization errors from the heavy-quark action
with power counting~\cite{Kronfeld:2000ck}.
We compute both the coefficient of the dimension 5 operator in the Fermilab action $c_{SW}$ and the rotation
parameter $d_1$ for the heavy-quark fields at tree level in tadpole-improved lattice perturbation
theory~\cite{ElKhadra:1996mp}.
Then the leading heavy-quark errors in $h_+(1)$ are of $\mathrm{O}\left(\alpha_s(\Lambda/2m_Q)^2\right)$ 
and $\mathrm{O}\left((\Lambda / 2m_Q)^3\right)$, where $\Lambda$ is a typical hadronic scale.
Using the values $\alpha_s = 0.3$, $\Lambda = 500$~MeV, and $m_c = 1.2$~GeV, we estimate that heavy-quark
discretization errors in $h_+(1)$ are $\sim$1--2\%.
At nonzero recoil, $w>1$, there are corrections to $h_+(w)$ of $\mathrm{O}\left(\alpha_s\Lambda/2m_Q\right)$,
but these are suppressed by $(1-w)$ because they vanish in the limit $w=1$ by Luke's
theorem~\cite{Luke:1990eg}.
We expect them to be largest at our highest recoil point $w=1.2$, and estimate their size to be $\sim$ 1\%.
Thus we estimate the uncertainty in $h_+(w)$ from heavy-quark discretization errors to be 2\%, which leads to
negligible errors in $R(D)$ and $P_L(D)$.
The leading heavy-quark error in $h_-(w)$ is of $\mathrm{O}\left(\alpha_s\Lambda/2m_Q\right)$, which
we estimate with the input parameters above to be $\sim$~6\%.
To be conservative, we take the error in the ratio $h_-(w)/h_+(w)$ to be 10\%, which leads to small errors in $R(D)$
and $P_L(D)$.

Our methods for computing $B\to D$ transitions incorporate the bulk of the matching of the lattice
vector current to continuum automatically, leaving a factor $\rho_{V_{cb}^\mu}$ close
to~unity~\cite{Harada:2001fj}.
For $R(D)$ and $P_L(D)$, only the relative matching of the spatial and temporal components of 
the current matters, $\rho_{V_{cb}^i}/\rho_{V_{cb}^{0}}$.
Although we have a one-loop calculation of $\rho_{V_{cb}^{0}}$ in hand, no nontrivial result for 
$\rho_{V_{cb}^{i}}$ is available.
We take $\rho_{V_{cb}^i}/\rho_{V_{cb}^{0}}=1.0\pm0.2$ to estimate the uncertainty from this source.
In similar calculations, we have never seen $\rho$~factors that differ from unity by more than 5\%, so this 
range is extremely conservative.
The uncertainty in the current renormalization factors leads to a small error in $R(D)$, but is the
second-largest source of error in $P_L(D)$, after statistics.

We also consider the systematic uncertainties from tuning the light-quark masses and determining the absolute
lattice scale $r_1$, but these produce negligible errors in both $R(D)$ and $P_L(D)$.

\begin{table}[tb]
    \caption{Error budgets for the branching fraction and longitudinal polarization ratios discussed in the 
        text. 
        Errors are given as percentages.}
    \begin{tabular}{lcccc}
    \hline\hline
    Source & $R(D)$ & $P_L(D)$  \\
    \hline
          Monte-Carlo statistics  & 3.7  &  1.2  \\
          Chiral-continuum extrapolation & 1.4 &  0.1 \\
           $z$-expansion & 1.5  &  0.1 \\
          Heavy-quark mass ($\kappa$) tuning  &  0.7  &0.1 \\
          Heavy-quark discretization & 0.2  &  0.3 \\
          Current $\rho_{V_{cb}^i}/\rho_{V_{cb}^{0}}$ & 0.4 & 0.7  \\
    \hline
           total  &  4.3\%  &  1.5\%\\
    \hline\hline
    \end{tabular}
\label{tab:ErrBud}
\end{table}


{\it Results and Conclusions}.
-- We obtain the following determinations for the branching-fraction and longitudinal-polarization ratios for
$B\to D\ell\nu$ semileptonic decay:
\begin{eqnarray}
	R(D)   &=& 0.316(12)(7) \,, \label{eq:RD-num} \\
	P_L(D) &=& 0.325(4)(3)  \,, \label{eq:PL-num}
\end{eqnarray}
where the errors are statistical and total systematic, respectively.
The value of $R(D)$ is approximately 1$\sigma$ larger than the recent Standard-Model
values obtained using estimates of $f_0(q^2)$ from Refs.~\cite{Kamenik:2008tj,Fajfer:2012vx}, 
but it is still 1.7$\sigma$ lower than the recent BaBar measurement,
$R(D) = 0.440 \pm 0.058 \pm 0.042$~\cite{:2012xj}.
The results for $R(D)$ from Belle~\cite{Adachi:2009qg} agree with those of BaBar, but have larger
uncertainties.
Current experimental measurements of $R(D)$ are statistics-limited, so the luminosities available at Belle~II
and SuperB should enable significant improvement on $R(D)$ and possibly a determination of $P_L(D)$.

\begin{figure}[b]
    \includegraphics[width=\linewidth]{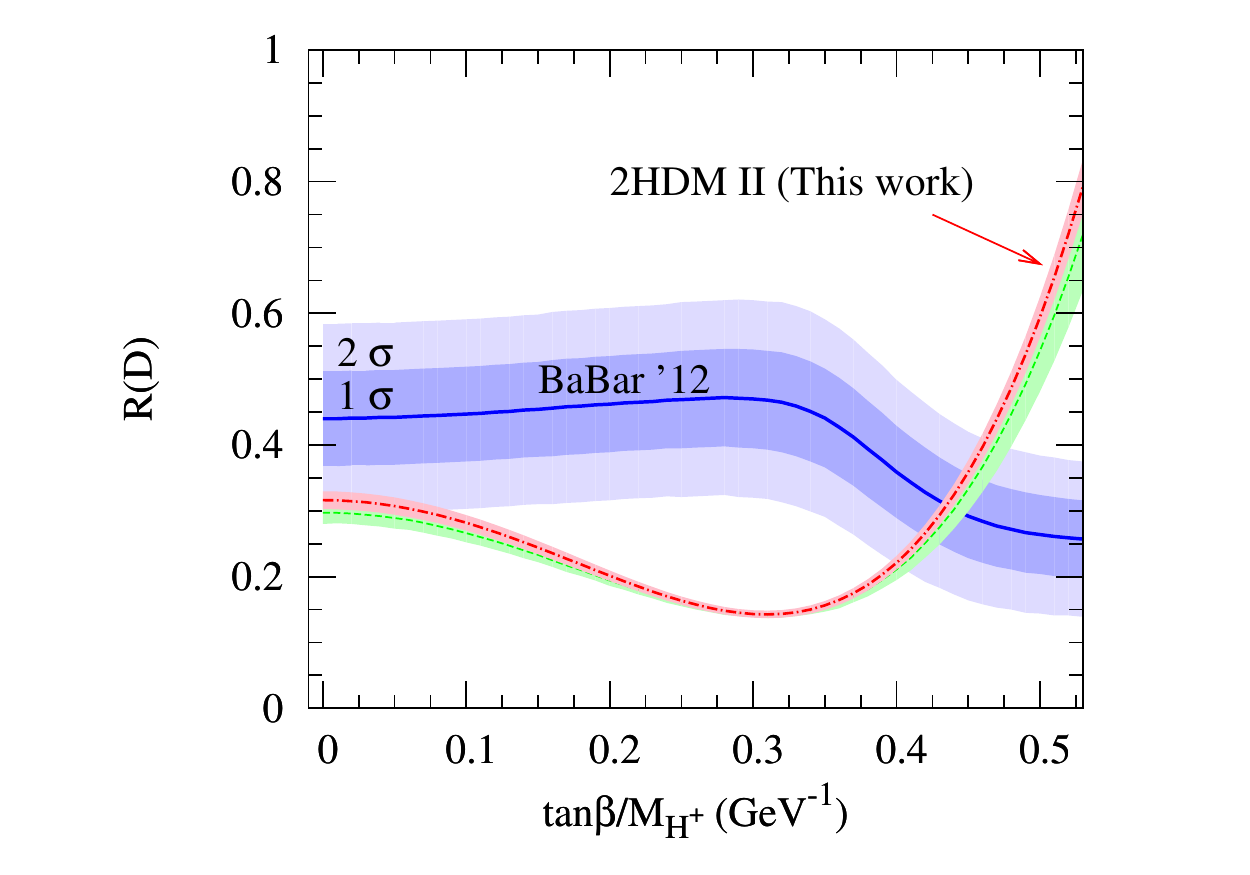}
    \caption{Lattice-QCD calculation (red) and experimental measurement~\cite{:2012xj} (blue) of $R(D)$ 
    vs.~$\tan\beta/M_{H^\pm}$ in the 2HDM~II.
    For the BaBar result, the dark- (light-) blue regions denote the 1$\sigma$ (2$\sigma$) error bands.
    For comparison, the 2HDM~II curve based on Refs.~\cite{Kamenik:2008tj,Fajfer:2012vx} and used in 
    Ref.~\cite{:2012xj} is shown in green.}
    \label{fig:rd}
\end{figure}

We also re-examine the interpretation of the BaBar measurement of $R(D)$ as a constraint on the 2HDM~II; the
result is plotted in Fig.~\ref{fig:rd}.
For this theory, the scalar-exchange coupling in Eq.~(\ref{eq:dGammaPlus}) is given by
\begin{equation}
    G_S^{\ell cb}  = G_F V_{cb} \frac{m_\ell(m_c + m_b\tan^2\beta)}{M_{H^\pm}^2}.
    \label{eq:GS}
\end{equation}
As soon as $\tan\beta\gtrsim4$, $m_c$ in Eq.~(\ref{eq:GS}) can be neglected.
On the other hand, evaluating Eq.~(\ref{eq:dGammaPlus}) depends sensitively on $m_c/m_b$ via the prefactor 
multiplying $f_0(q^2)$.
We take $m_c/m_b=0.22$, following Refs.~\cite{Fajfer:2012vx,:2012xj}.
Our improved calculation of the scalar form factor $f_0(q^2)$ increases the prediction for $R(D)$ by about 
$1\sigma$ for all $\tan\beta/M_{H^+}<0.6~\text{GeV}^{-1}$.
This sensitivity of $R(D)$ to differences in $f_0(q^2)$ suggests that one
should be cautious in using indirect estimates of the form factors to constrain new-physics models in other
decay channels such as $B \to D^*\tau\nu$. 

Inspection of the general formulas for the differential decay rates, Eqs.~(\ref{eq:dGammaMinus})
and~(\ref{eq:dGammaPlus}), shows that many new-physics explanations for the $\sim 3\sigma$ tension
between measurements of $\{R(D), R(D^*) \}$ and the Standard Model are possible.
Even in the context of 2HDM, variants other than type~II can have different
phenomenology~\cite{Crivellin:2012ye}.
Lattice-QCD calculations of $f_+(q^2)$ and $f_0(q^2)$ can be used to provide reliable predictions for $R(D)$
in any model given values for the couplings $\{ G_S, G_V, G_T\}$.
We note, however, that the Dalitz distribution for the lepton and $D$-meson energies in the $B$-meson rest
frame may be more sensitive to tensor interactions than $R(D)$~\cite{Kronfeld:2008gu}.

The largest source of uncertainty in our determinations of $R(D)$ and $P_L(D)$ is statistical errors, which
we expect to reduce with an analysis of our full data set in a future work~\cite{Qiu:2011ur}.
The ratio $R(D)$ is correlated with $P_L(D)$ as well as other observables such as $R(D^*)$ or $R(D_s)$ in
many new-physics models; thus the pattern of experimental results for these quantities can help
distinguish between new-physics scenarios, such as those with and without a charged-Higgs
boson~\cite{Tanaka:2010se}.
We will present lattice-QCD results for $R(D^*)$ and $P_L(D^*)$ in a future paper on the $B \to D^* \ell \nu$
form factors, and also note that we could easily obtain Standard-Model predictions for $R(D_s)$ and
$P_L(D_s)$ if measurements of these quantities were possible with an $\Upsilon(5S)$ run at a $B$~factory.

Given the present tensions between experimental measurements and Standard-Model
predictions for both $\{R(D), R(D^*) \}$ and the leptonic branching fraction $\text{BR}(B \to \tau
\nu)$~\cite{Bona:2009cj,Lunghi:2010gv,Charles:2011va,Laiho:2012ss}, lattice-QCD calculations of $B\to D\tau
\nu$ form factors and other hadronic weak matrix elements can play a key role in revealing whatever theory
beyond the Standard Model is realized in Nature.


{\it Acknowledgements}. -- We thank Manuel Franco Sevilla, Jernej Kamenik, and Bob Kowaleski for valuable discussions.
We thank Yang Bai for emphasizing the importance of the choice of scale for $m_c$ in Eq.~(\ref{eq:dGammaPlus}).
Computations for this work were carried out with resources provided by 
the USQCD Collaboration, the Argonne Leadership Computing Facility, 
the National Energy Research Scientific Computing Center, and the Los Alamos National Laboratory, which
are funded by the Office of Science of the United States Department of Energy; and with resources
provided by the National Institute for Computational Science, the Pittsburgh Supercomputer Center, 
the San Diego Supercomputer Center, and the Texas Advanced Computing Center, 
which are funded through the National Science Foundation's Teragrid/XSEDE Program. 
This work was supported in part by the U.S. Department of Energy under Grants 
No.~DE-FG02-91ER40628 (C.B.), 
No.~DOE~FG02-91ER40664 (Y.M.),
No.~DE-FC02-06ER41446 (C.D., J.F., L.L., M.B.O.), 
No.~DE-FG02-91ER40661 (S.G., R.Z.), 
No.~DE-FG02-91ER40677 (D.D., A.X.K.), 
No.~DE-FG02-04ER-41298 (J.K., D.T.); 
by the National Science Foundation under Grants 
No.~PHY-1067881, No.~PHY-0757333, No.~PHY-0703296 (C.D., J.F., L.L., M.B.O.), 
No.~PHY-0757035 (R.S.); 
by the Science and Technology Facilities Council and the Scottish Universities Physics Alliance (J.L.);
by the MICINN (Spain) under grant FPA2010-16696 and Ram\'on y Cajal program (E.G.);
by the Junta de Andaluc\'ia (Spain) under Grants No.~FQM-101, No.~FQM-330, and No.~FQM-6552 (E.G.);
by European Commission (EC) under Grant No.~PCIG10-GA-2011-303781 (E.G.);
and by the Creative Research Initiatives program (3348-20090015) of the NRF grant funded by the Korean 
government (MEST) (J.A.B.).
This manuscript has been co-authored by employees of Brookhaven Science Associates, LLC,
under Contract No.~DE-AC02-98CH10886 with the U.S. Department of Energy. 
Fermilab is operated by Fermi Research Alliance, LLC, under Contract No.~DE-AC02-07CH11359 with
the U.S. Department of Energy.

\bibliography{BtoDTau}
\bibliographystyle{apsrev4-1} 

\end{document}